\documentclass[prb,twocolumn]{revtex4-1}

\usepackage{graphicx}
\usepackage{amsmath}
\DeclareMathOperator{\csch}{csch}

\usepackage{color}
\usepackage{amssymb}
\usepackage{amstext}
\usepackage{amsthm}
\usepackage{amsfonts}
\usepackage{latexsym}
\usepackage{hyperref}
\hypersetup{
    colorlinks=true,
    linkcolor=blue,
    urlcolor=cyan,
}
\def\beq{\begin{equation}}{\it}
\def\eeq{\end{equation}}
\def\beqa{\begin{eqnarray}}{\it}
\def\eeqa{\end{eqnarray}}

\newcommand{\bra}[1]{{ \left\langle #1 \right|} }
\newcommand{\ket}[1]{{ \left|  #1 \right\rangle } }

\begin{document}

\title{The Hellmann-Feynman theorem at finite temperature}

\author{Marina   Pons }
\affiliation{Departament de F\'isica Qu\`antica i Astrof\'isica,\\
Facultat de F\'{\i}sica, Universitat de Barcelona, E--08028 Barcelona, Spain}

\author{Bruno Juli\'a-D\'iaz}
\affiliation{Departament de F\'isica Qu\`antica i Astrof\'isica,\\
Facultat de F\'{\i}sica, Universitat de Barcelona, E--08028 Barcelona, Spain}
\affiliation{Institut de Ci\`encies del Cosmos, 
Universitat de Barcelona, IEEC-UB, Mart\'i i Franqu\`es 1, E--08028
Barcelona, Spain}

\author{Arnau  Rios}
\affiliation{Department of Physics, Faculty of Engineering and Physical Sciences,\\
University of Surrey, Guildford, Surrey GU2 7XH, United Kingdom} 

\author{Isaac  Vida\~na } 
\affiliation{Istituto Nazionale di Fisica Nucleare, Sezione di Catania. Dipartimento di Fisica e Astronomia ``Ettore Majorana", Universit\`a di Catania, Via Santa Sofia 64, I-95123 Catania, Italy}

\author{Artur Polls}
\affiliation{Departament de F\'isica Qu\`antica i Astrof\'isica,\\
Facultat de F\'{\i}sica, Universitat de Barcelona, E--08028 Barcelona, Spain}
\address{Institut de Ci\`encies del Cosmos, 
Universitat de Barcelona, IEEC-UB, Mart\'i i Franqu\`es 1, E--08028
Barcelona, Spain}

%%%%%%%%%%%%%%%%%%%%%%%%%%%%%%%%%%%%%%%%%%%%%%%%%
\begin{abstract}
We present a simple derivation of the Hellmann-Feynman theorem at 
finite temperature. We illustrate its validity by considering three 
relevant examples which can be used in quantum mechanics lectures: 
the one-dimensional harmonic oscillator, the one-dimensional Ising model 
and the Lipkin model. We show that the Hellmann-Feynman theorem allows one 
to calculate expectation values of operators that appear in the Hamiltonian. This is 
particularly useful when the total 
free energy is available, but there is no direct access to the thermal average of the operators themselves.  
\end{abstract}
%%%%%%%%%%%%%%%%%%%%%%%%%%%%%%%%%%%%%%%%%%%%%%%%%
\maketitle
%%%%%%%%%%%%%%%%%%%%%%%%%%%%%%%%%%%%%%%%%%%%%%%%%

\section{Introduction} 
\label{intro}

The Hellmann-Feynman (HF) theorem~\cite{hellmann1933,feynman1939} at 
zero temperature is a very useful tool in quantum mechanics. 
Its most important application involves
the calculation of expectation values of operators contained in the Hamiltonian. 
The theorem
is often introduced in undergraduate quantum mechanics 
courses,~\cite{griffiths1995,atkins2011} and has been derived and generalised 
in different ways. 
The HF is closely connected to the virial theorem.~\cite{Dongpei1986,pekolu2016}
In the context of perturbation theory, Epstein showed that the HF theorem 
provides a consistent picture of wavefunction renormalization. \cite{Epstein1954}
It was later used to generate Rayleigh-Schr{\"o}dinger perturbation theory corrections. \cite{Singh1989,Almeida2000}
In its simplest version, the theorem is used for 
non-degenerate states. Generalizations
to degenerate subspaces also exist.~\cite{Singh1989,alon2003} 
The theorem is versatile and has been extended to off-diagonal expectation values;~\cite{Klein1978,Castro1980}
to Gamow states; \cite{Ziesche1987}
to linear superpositions of energy eigenstates \cite{Balasubramanian1994} and
to cases where the domain of definition of the Hamiltonian depends on a parameter. \cite{Esteve2010}
Analogues of the theorem can be found in classical systems, \cite{McKinley1971} and  a time-dependent extension has been formulated too. \cite{Hayes1965,Epstein1966,Sun2016}

Applications of the theorem span a large number of sub-fields. 
In atomic physics, the HF theorem can be used to find 
closed expressions for useful expectation values. \cite{SanchezdelRio1982,Valk1986}
In quantum chemistry, the HF theorem 
has been used to evaluate molecular 
forces and to calculate Coulomb interaction energies.~\cite{fitts2002,jensen2007}
The quark mass dependence in some hadronic systems can be accessed through
the HF theorem.  \cite{Quigg1979,Lichtenberg1989}
Efficient implementations of the HF theorem in diffusion 
Monte Carlo simulations provide direct access to kinetic and potential energies
of bosonic systems. \cite{Vitiello2011}
In nuclear physics, the theorem has been used recently 
to extract contributions of the nuclear force 
to the nuclear symmetry \cite{vidana11}
and spin symmetry energies. \cite{vidana16} 

In contrast to these zero-temperature cases, the use of 
the HF theorem at finite temperature is much more scarce. 
The first derivation that we are aware of is provided in Ref.~\onlinecite{cabrera1989}. 
This was followed by independent derivations in Refs.~\onlinecite{fan1995}, 
\onlinecite{Popov1998} and \onlinecite{rai2007} each with a somewhat different focus.
When it comes to actual applications, 
the HF theorem is rarely employed at finite temperature.
Exceptions include 
the calculation of correlation functions in 1D Bose gases~\cite{Kheruntsyan2003} 
and  the evaluation of specific matrix elements in lattice QCD.~\cite{lattice2017}
In heavy-ion physics,
the HF theorem provides a direct link between thermodynamical consistency
and quasi-particle descriptions. \cite{Shanenko2003}
In the context of strongly correlated fermionic systems, the adiabatic sweep relations 
derived by Tan \cite{Tan2008} arise naturally in a HF formulation, 
and have been used to study finite temperature systems in Ref.~\onlinecite{Hu2011}.
The theorem has been used to link thermodynamical properties to microscopic
density matrices in the context of quantum phase transitions. \cite{Wei2018}

The scope of the present paper is to present a simple derivation of the 
HF theorem at finite temperature, as a useful resource in teaching quantum
mechanics at finite temperature. 
We complement the derivation of the theorem with three 
illustrative, and relatively different, quantum mechanical examples: 
the harmonic oscillator, the Ising 
model~\cite{ising1925} and the Lipkin model.~\cite{lipkin1965}
Some of these examples are analytical,  
and can be used in undergraduate courses to enlighten 
the meaning and relevance of the theorem. 
Along the way, we will answer
natural questions about the suitable thermodynamical 
potentials that are necessary for the derivation of the HF theorem. 
Should one use, for instance, the energy or the free energy? 
Our approach will also help us identify the role of the entropy in the HF theorem 
at finite temperature.

%%%%%%%%%%%%%%%%%%%%%%%%%%%%%%%%%%%%%%%%%%%%%%%%%
\section{The Hellmann-Feynman theorem}

\subsection{Zero temperature}

The HF theorem allows one to determine the expectation 
value of an operator contained in the Hamiltonian for a given 
non-degenerate eigenstate. To derive the theorem, 
one defines a parametric, $\lambda$-dependent, Hamiltonian, 
$H^\lambda$. In the examples presented in this paper, we
make use of a linear dependence in $\lambda$, 
\begin{align}
H^\lambda = H_0 + \lambda H_1\, , \label{eq:linearl}
\end{align}
but this does not need to be the case. 
For $\lambda=1$, the Hamiltonian in Eq.(\ref{eq:linearl}) is automatically decomposed in 
two pieces, $H = H_0 + H_1,$ where $H_1$ is the operator for which we want 
to calculate the expectation value.
Typically, these operators are the kinetic or 
the potential energies but, as we shall see in the following,
other options arise depending on the model and the corresponding Hamiltonian.

The $n^{th}$-eigenvalue of $H^\lambda$, $E_n^\lambda$, is the solution of the eigenvalue problem
\begin{equation}
	H^\lambda \ket{ \psi_n^\lambda } = 
	E_n^\lambda \ket{ \psi_n^\lambda } \,. 
\label{eq:1}
\end{equation}
Hereafter, all observables denoted with a $\lambda$ superscript are to be understood as expectation values over 
$\lambda$-dependent states, $\ket{ \psi_n^\lambda }$.
For a linear parametric dependence on $\lambda$, Eq.~(\ref{eq:linearl}), 
the zero-temperature HF theorem states that one can evaluate 
the expectation value 
$\bra{ \psi_n } H_1 \ket{ \psi_n}$ from the derivative with respect to $\lambda$ of the corresponding eigenenergies,
\begin{equation}
\bra{ \psi_n } H_1 \ket{ \psi_n} = 
\left. \frac {d E_n^\lambda}{d \lambda }\right|_{\lambda = 1} \,.
\label{eq:2}
\end{equation}
The theorem is useful because a direct calculation
of the expectation value on the left hand side is sometimes more cumbersome than 
the direct evaluation of the energy derivative on the right hand side.

We now proceed to prove this zero-temperature result, as this sets the ground for the
finite temperature discussion. The proof has
been discussed in many standard quantum 
mechanics books,~\cite{griffiths1995,atkins2011} and it stems from a careful analysis of the
derivative of the eigenvalue defined by Eq.~(\ref{eq:1}). The derivative is in fact decomposed in
three terms,
\begin{align}
\frac {d E_n^\lambda}{d \lambda } &=
\frac {d}{d\lambda}\left [ \bra{ \psi_n^\lambda } H^\lambda
\ket{ \psi_n^\lambda }  \right ] \nonumber \\
&= \left ( \frac {d}{d \lambda} \bra{ \psi_n^\lambda } \right )
 H^\lambda \ket{ \psi_n^\lambda} \nonumber 
+
\bra{ \psi_n^\lambda }  \left( \frac {d}{d \lambda} H^\lambda \right)
\ket{ \psi_n^\lambda } \\
&+
\bra{ \psi_n^\lambda } H^\lambda 
\left ( \frac {d}{d \lambda} \ket{ \psi_n^\lambda } \right) \, .
\label{eq:3}
\end{align}
Taking into account that $\ket{ \psi_n^\lambda }$ is an 
eigenvector of $H^\lambda$, the first and third terms can 
be combined to yield
\beqa
\frac {d E_n^\lambda}{d \lambda }&=&  
E_n^\lambda \frac {d}{d \lambda} 
\left [ \langle \psi_n^\lambda \mid  \psi_n^\lambda \rangle \right ]
	+ \langle \psi_n^\lambda \mid  \frac {d H^\lambda}{d \lambda} 
	\mid \psi_n^\lambda \rangle \,.
\label{eq:4}
\eeqa
The eigenstates are normalized, and therefore the derivative with respect 
to $\lambda$ of the overlap 
$\langle \psi_n^\lambda\mid  \psi_n^\lambda \rangle$ is zero and we recover
\beqa
\frac {d E_n^\lambda}{d \lambda }&=&  
 \langle \psi_n^\lambda \mid  \frac {d H^\lambda}{d \lambda} 
	\mid \psi_n^\lambda \rangle \,.
\label{eq:4b}
\eeqa
The HF theorem, Eq.~(\ref{eq:2}), is obtained for a linear parametric dependence, Eq.~(\ref{eq:linearl}), and for $\lambda=1$. In particular, we note that the 
relation holds for any state, $n$, and is true for the ground state, $n=0$.

\subsection{Finite temperature}

The HF theorem at finite temperature has been much 
less discussed in the literature.~\cite{cabrera1989,fan1995,Popov1998,rai2007} 
In the following, we employ the canonical ensemble in our derivation.
Starting, again, from 
a $\lambda-$dependent parametric Hamiltonian, we define a $\lambda-$dependent 
density matrix operator,
\begin{equation}
 \rho^{\lambda} = e^{- \beta H^\lambda} \, ,
	\label{eq:5}
\end{equation}
where $\beta = 1 /T$ is the inverse of the temperature. The 
partition function $Z^{\lambda}$ is defined as the trace of 
$\rho^{\lambda}$, and can be evaluated in the basis of 
eigenvectors of $H^{\lambda}$,
\begin{equation}
	Z^\lambda = \text{Tr} \rho^{\lambda} = \sum_n e^{- \beta E_n^\lambda} \,.
\label{eq:6}
\end{equation}
From the partition function, one can evaluate relevant thermodynamical potentials like
the average thermal energy,
\begin{equation}
E^\lambda = -\frac {\partial}{\partial \beta } \ln Z^\lambda \, .
\label{eq:ener}
\end{equation}
Further, we note that the free energy is given by the expression
\begin{equation}
	F^\lambda = - \frac{1}{\beta} \ln {Z^\lambda} = 
	- T \ln {\sum_n e^{ - \beta E_n^\lambda}}\,, 
	\label{eq:7}
\end{equation}
and the entropy, in turn, is given by
\begin{equation}
S^\lambda= -\frac{\partial F^\lambda}{\partial T} \, . 
\label{eq:entro}
\end{equation}
The 
derivative  of $F^\lambda$  with respect to $\lambda$ yields
\beqa
\frac {\partial F^\lambda}{\partial \lambda} &=& 
\frac {1}{Z^\lambda } 
\sum_n e^{-\beta E_n^\lambda}~ \frac {d E_n^\lambda } {d \lambda} \, .
\eeqa
The eigenvalue derivatives, $d E_n^\lambda/d \lambda$, are
independent of the temperature and given by
Eq.~(\ref{eq:4b}).
As a consequence, the derivative of the free energy can be expressed as a
thermal ensemble average of $\partial H^\lambda / \partial \lambda$,
\beqa
	\frac {\partial F^\lambda}{\partial \lambda } &=& 
	\frac {1}{Z^\lambda} \sum_n e^{- \beta E_n^\lambda} 
	\bra{ \psi_n^\lambda }
	\frac {\partial H^\lambda}{\partial \lambda} 
	\ket{ \psi_n^\lambda } \nonumber \\
	&=&
	\left\langle \frac{\partial H^\lambda}{\partial \lambda} \right\rangle_T ,
\eeqa
at a given value of $\lambda$. In particular, for a linear parametric dependence, Eq.~(\ref{eq:linearl}), and for $\lambda= 1$, one recovers 
the thermal average of $H_1$ at a temperature $T$, $\langle H_1 \rangle_T$:
\beqa
\left. \frac {\partial F^\lambda}{\partial \lambda}\right|_{\lambda=1} 
= \frac {1}{Z} \sum_n e^{-\beta E_n} 
\bra{ \psi_n }  H_1 \ket{ \psi_n } = 
\langle H_1 \rangle_T \,.
\label{eq:HFt}
\eeqa
Note that this expression is very similar to the HF
theorem at $T=0$, Eq.~(\ref{eq:2}), 
but replacing the energy, $E^\lambda$, with the free energy, $F^\lambda$. 
This is the main result of this paper, which we take as the generalization of the HF theorem to
finite temperature. We will use this result in the following sections to 
illustrate the usefulness of the theorem.

An additional useful result can be derived as a corollary.
At finite temperature, the energy and the free energy differ by a factor proportional to the entropy.
At arbitrary $\lambda$, the difference is given by the expression
\beqa
F^\lambda = E^\lambda - T S^\lambda \,.
\eeqa
This relation suggests that one can also find a HF-like expression for the entropy. In fact, taking
the derivative with respect to $\lambda$ of Eq.~(\ref{eq:entro}), one finds
\beqa
\frac{ \partial S^\lambda}{\partial \lambda}= - \frac{ \partial^2 F^\lambda}{\partial \lambda \partial T}\,.
\eeqa
Switching the order of the derivatives and using Eq.~(\ref{eq:HFt}), we obtain
\beqa
\left.\frac {\partial S^\lambda}{\partial \lambda} \right|_{\lambda = 1}  
= - \frac {\partial \langle H_1 \rangle_T }{\partial T } \,.
\label{eq:HFs}
\eeqa
This expression suggests that the entropy of the system is directly connected to
the temperature derivatives of thermal expectation values of $H_1$. This is 
to a certain extent unexpected, because the entropy itself is not directly related to thermal
averages of parts of the Hamiltonians.

In the following three sections, we illustrate the HF theorem with three different
Hamiltonians and settings that are often  discussed in quantum-mechanics teaching. 
Each example shows a different application of the theorem to compute different
contributions to the system's energies or thermodynamical potentials.

%%%%%%%%%%%%%%%%%%%%%%%%%%%%%%%%%%%%%%%%%%%%%%%%%

\section{Harmonic oscillator}

We start with the textbook 
example of a one-dimensional harmonic oscillator (h.o.). A unique advantage of this example is
that it can be treated analytically at finite temperature. We use h.o.~units, in which the Hamiltonian
reads,
\beqa
H = - \frac {1}{2}\frac {d^2}{d x^2} + \frac {1}{2} x^2 \,.
\eeqa
The eigenvalues are $E_n = \left( n+ \frac{1}{2} \right)$ for $n=0,1,2,\cdots$.

The
application of the HF theorem at zero temperature is straightforward, and allows 
a direct evaluation of the h.o.~potential term. One  
defines the $\lambda-$dependent Hamiltonian
\beqa
H^\lambda = - \frac {1}{2}\frac {d^2}{d x^2} + \frac {\lambda}{2} x^2 \, ,
\eeqa
so that $H_0$ is the kinetic term and $H_1$ is the h.o.~potential.
In practice, $\lambda$ renormalizes the h.o.~frequency by a factor $\sqrt{\lambda}$, and 
consequently the eigenvalues of $H^\lambda$ are given by the expression 
$E_n^\lambda= \left( n+ \frac{1}{2} \right) {\sqrt{\lambda}}$. At zero temperature, the 
HF theorem is useful in computing the expectation 
value of the harmonic potential in any eigenstate:
\begin{align}
\bra{ \psi_n } \frac{x^2}{2} \ket{ \psi_n}&=
\left. \frac {\partial E_n^\lambda}{\partial \lambda} \right|_{\lambda=1} 
= \left.\left (n+ \frac {1}{2}\right ) \frac {1}{2 \sqrt{\lambda}} 
\right|_{\lambda=1} \nonumber \\
&= \frac {1}{2} \left (n + \frac{1}{2} \right ) \,.
\end{align}
This expression is in agreement with the virial theorem. 

At finite temperature, the expressions for the partition function associated to $H$
can be found in statistical mechanics textbooks,~\cite{pathria}
\beqa
Z = \sum_{n=0}^{\infty}  e^{- \beta \left(n+\frac{1}{2} \right)} = 
\frac {1}{2} \csch {\left ( \frac {\beta}{2} \right )} \,. 
\eeqa
From the partition function, one can derive analytical expressions for the free energy
\beqa
F = -\frac {1}{\beta} \ln {Z}= \frac {1}{2} + \frac {1}{\beta} 
\ln {( 1 - e^{- \beta} )}\,;
\eeqa
the average thermal energy,
\beqa
E = - \frac {\partial }{\partial \beta} \ln {Z} = \frac {1}{2} + 
\frac {e^{- \beta}}{1 -e^{-\beta}} \,;
\label{eq:ehoT}
\eeqa
and  the entropy, 
\beqa
S = - \ln (1 - e^{- \beta}) + \frac { \beta e^{-\beta}}{1 - e^{-\beta}} \,.
\eeqa

Similar expressions hold for the thermodynamical potentials associated to the parametric
Hamiltonian $H^\lambda$. The associated free energy is given by,
\beqa
F^\lambda = \frac {\sqrt {\lambda}}{2} + \frac {1}{\beta} \ln {(1 - e^{-\beta \sqrt {\lambda}})} \,.
\eeqa
This expression can then be used in the finite temperature HF theorem, Eq.~(\ref{eq:HFt}),  
to find the thermal average of the harmonic 
potential:
\beqa
\left\langle \frac{x^2}{2} \right\rangle_T = 
\left. \frac {\partial F^\lambda}{\partial \lambda}\right|_{\lambda=1}
=\frac {1}{4} + \frac {1}{2} \frac { e^{-\beta}}{1 - e^{-\beta} } \,. 
\label{eq:avT}
\eeqa
This result agrees with the finite temperature virial theorem,
which states that the potential and the kinetic energies each contribute
a half of the total energy, Eq.~(\ref{eq:ehoT}):
\beqa
\left\langle \frac{x^2}{2} \right\rangle_T = 
\frac{1}{2} E \, .
\label{eq:vir}
\eeqa

\begin{figure}[t!]
\centering
 \includegraphics[width=0.75\columnwidth]{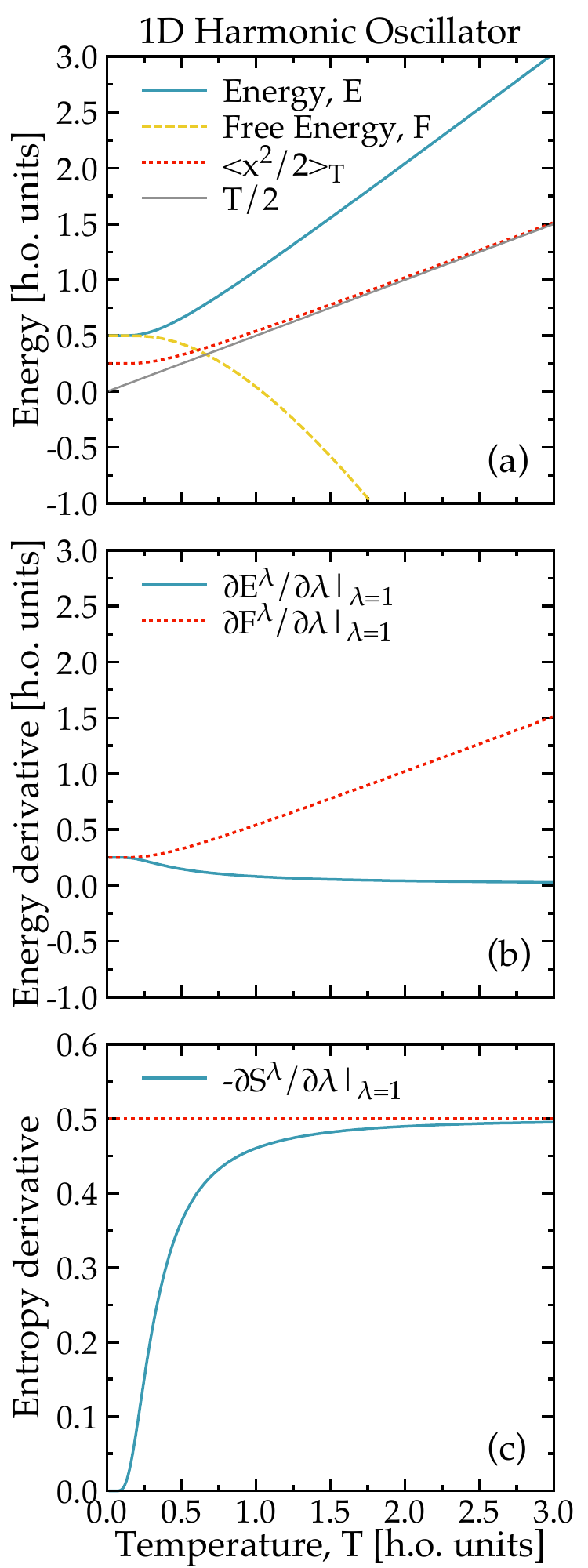}
\caption{(a) Energy $E$ (solid line), free energy $F$ (dashed line) 
and $\left\langle {x^2}/{2} \right\rangle_T$ (dotted line) as a 
        function of temperature for a one dimensional harmonic oscillator. The solid line, $E=\frac{T}{2}$, illustrates the classical limit of the potential energy. 
(b) Derivative of $E^\lambda$ (solid line) and $F^\lambda$ (dotted line) with respect to $\lambda$ at $\lambda=1$ as a function of temperature. 
(c) Minus the derivative of $S^\lambda$  with respect to $\lambda$ at $\lambda=1$ 
as a function of temperature (solid line). The sign is chosen so this expression coincides with 
$\frac {\partial }{\partial T } \left\langle \frac{x^2}{2} \right\rangle_T$. The dotted line corresponds
to the asymptotic value $\left. {\partial S^\lambda}/{\partial \lambda} \right|_{\lambda = 1} \to -1/2 $.
}
\label{fig:ho}
\end{figure}

Panel (a) of Fig.~\ref{fig:ho} shows the thermodynamical potentials $E$ (solid line) and $F$ (dashed line), as well
as the thermal expectation value of ${x^2}/{2}$ (dotted line) for 
a one dimensional harmonic oscillator as a function of temperature. 
As expected, both $E$ and $F$ tend to $1/2$ at very low temperatures. In contrast,
for large $T$, classical statistics dictates that $E \xrightarrow[T\gg1]{} T$ and 
 $F \xrightarrow[T\gg1]{} -T \ln T$. For all values of temperature, 
 we find that Eq.~(\ref{eq:vir}) holds, in agreement with the virial theorem.

Further, we report in panel (b) of Fig.~\ref{fig:ho} the two derivatives
        $\left. {\partial E^\lambda}/{\partial \lambda} \right|_{\lambda = 1} $ (solid line) and
                $\left. {\partial F^\lambda}/{\partial \lambda} \right|_{\lambda = 1} $ (dashed line)
as a function of the temperature. Notice that 
$\left. {\partial F^\lambda}/{\partial \lambda} \right|_{\lambda = 1} $
coincides with  
the thermal average  $\left\langle {x^2}/{2} \right\rangle_T$ shown in panel (a). 
In contrast, the corresponding derivative of the energy is not related to a thermal average of the Hamiltonian. 
The asymptotic behaviours of these derivatives are known. When $T \rightarrow 0$, 
both derivatives tend to the same value, $\left. {\partial E^\lambda}/{\partial \lambda} \right|_{\lambda = 1}  = \left.{\partial F^\lambda}/{\partial \lambda} \right|_{\lambda = 1}  \rightarrow 1/4$, 
which is precisely the expectation value of $x^2/2$ on the zero-temperature ground state, $n=0$. 
The classical limit for the two derivatives is however very different. For $T \rightarrow \infty$, 
$\left. {\partial E^\lambda}/{\partial \lambda} \right|_{\lambda = 1}   \rightarrow 0$ and is therefore negligible. 
The derivative of the free energy, however, is a linearly increasing function of $T$,
$\left. {\partial F^\lambda}/{\partial \lambda} \right|_{\lambda = 1}  \rightarrow {T}/{2}$.

%%%%%%%%%%%%%%%%%%%%%%%%%%%%%%%%%%%%
%%%%%%%%%%%%%%%%%%%%%%%%%%%%%%%%%%%%

The h.o.~case also allows for an explicit analytical test of
Eq.~(\ref{eq:HFs}).
On the one hand, 
the derivative of the thermal average of the harmonic potential 
can be explicitly calculated,
\beqa
\frac {\partial }{\partial T } \left\langle H_1\right\rangle_T=\frac {\partial }{\partial T } \left\langle \frac{x^2}{2} \right\rangle_T
= \frac {\beta^2}{2} \frac {e^{- \beta}}{(1-e^{-\beta })^2} \,.
\label{eq:derTpot}
\eeqa
On the other hand, $S^\lambda$ is written as, 
\beqa
S^\lambda = - \ln \left(1 - e^{-\beta \sqrt{\lambda}} \right) 
+ \sqrt{\lambda} \beta 
\frac {e^{- \beta \sqrt{\lambda}}}{1 - e^{\beta \sqrt{\lambda}}} \,,
\eeqa
and therefore,
\beqa
\left. \frac {\partial S^\lambda}{\partial \lambda} \right|_{\lambda=1} 
= -\beta^2 \frac {1}{2} \frac {e^{-\beta}}{(1-e^{-\beta})^2} \,,
\label{eq:26}
\eeqa
in agreement with Eq.~(\ref{eq:HFs}). 
This function is reported in the solid line of panel (c) of Fig.~\ref{fig:ho}. We note that 
this is a monotonously decreasing function of temperature  (notice the minus sign in the figure legend), 
which goes to zero 
at very low temperatures. In contrast, in the classical limit, the function asymptotically tends to $-\frac{1}{2}$ (dotted line). 
By virtue of Eq.~(\ref{eq:derTpot}), this implies that the classical limit of the
potential energy is ${T}/{2}$, as reported in panel (a).

%%%%%%%%%%%%%%%%%%%%%%%%%%%%%%%%%%%%%%%%%%%%%%%%%

\section{ Ising model}

The second example we consider is the Ising model,~\cite{ising1925}
which is also discussed in many textbooks of statistical mechanics~\cite{pathria} in connection 
with the study of ferromagnetic materials and phase transitions.
The HF theorem in this model can be used to estimate different terms of the Hamiltonian 
which have physical relevance. While analytical expressions for the partition
function are available, our discussion is based on numerical results for brevity. 

The Hamiltonian of the one-dimensional Ising model reads 
\begin{equation}
H = - J \sum_{\langle i,j \rangle} s_i s_j  - h \sum_i^N s_i \, .
\label{eq:i1}
\end{equation}
Here, $J$ is the coupling constant between two spins and the symbol 
$\langle i,j\rangle$ indicates that the spins interact only with their
nearest neighbours. $s_i$ are a 
collection of spin variables which can take values of  $\pm1$. 
We consider a one dimensional 
chain of $N$ spins with periodic boundary conditions. 
The Hamiltonian describes a ferromagnetic (antiferromagentic)
system for $J<0$ ($J>0$). $h$ defines the strength of an external (magnetic)
field. 

At zero temperature, the ground state corresponds to a configuration 
in which all the spins are pointing in the same direction. The finite temperature discussion is framed by 
the partition function, which in this case
and considering periodic  boundary conditions, 
is given by~\cite{pathria,baxter1982}
\begin{align}
Z =e^{N \beta J} & 
\left ( \left [ \cosh \beta h 
+ \sqrt { \sinh^2 \beta h + e^{- 4 \beta J}} \right ]^N \right . \nonumber \\
&+ \left .  \left [ \cosh \beta h 
- \sqrt { \sinh^2 \beta h + e^{- 4 \beta J }} \right ]^N \right ) \, .
\label{eq:i2}
\end{align}

\begin{figure}[t]
\centering
\includegraphics[width=0.75\columnwidth]{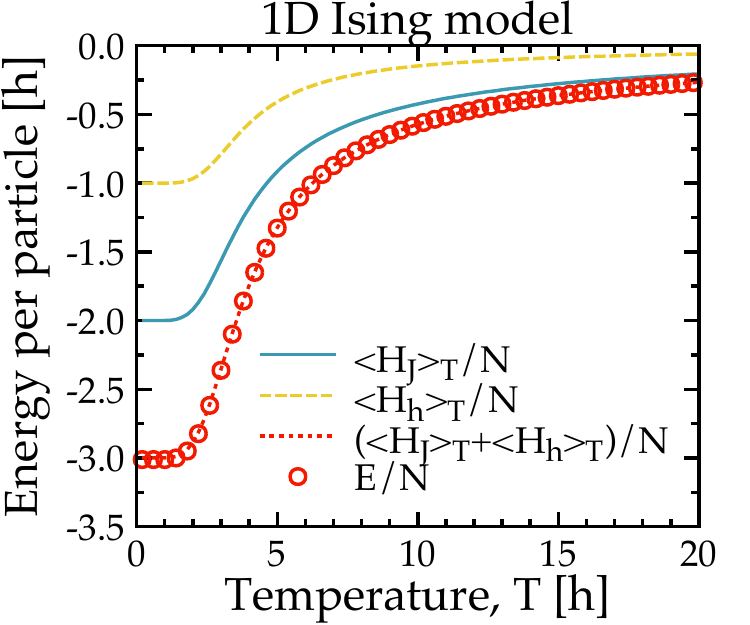}
\caption{ Thermal averages of $H_J$ (solid line), $H_h$ (dashed line) and $H$ (dotted line)
as a function of temperature for the Ising model with $J=2$ and $h=1$. Energies have been divided by the particle number ($N=10$). Energies and temperatures are measured in units of $h$.
\label{fig:ising}}
\end{figure}

The HF theorem gives access to the expectation values of either of the two terms in the
Hamiltonian. Let us first calculate the thermal average of 
the interaction term, so that $H_1 \equiv H_J = - J \sum s_i s_j $. We define
\begin{equation}
H^{\lambda_1} = -J \lambda_1 \sum_{\langle i,j\rangle} s_i s_j 
- h \sum_i s_i \,,
\label{eq:i5}
\end{equation}
which leads to the partition function $Z^{\lambda_1}$:
\begin{align}
Z^{\lambda_1} = e&^{N \beta \lambda_1 J}  \left ( 
\left [ \cosh \beta h + \sqrt { \sinh^2 \beta h 
+ e^{- 4 \lambda_1 \beta J}} \right ]^N \right . \nonumber \\
&+ \left .  \left [ \cosh \beta h 
- \sqrt { \sinh^2 \beta h + e^{- 4 \lambda_1 \beta J }} \right ]^N 
\right ) \, .
\label{eq:i6}
\end{align}
The corresponding free energy is readily computed from 
$F^{\lambda_1}=-\frac{1}{\beta} \ln Z^{\lambda_1}$. 
With this, the thermal average of the interaction energy is given by
\begin{equation}
\langle H_J \rangle_T =
 \left. \frac {\partial F^{\lambda_1}}{\partial \lambda_1} 
 \right|_{\lambda_1 = 1} \, .
\label{eq:i7}
\end{equation}
The temperature dependence of $\langle H_J \rangle_T/N$, 
calculated according to the HF theorem,
is reported in Fig.~\ref{fig:ising} (solid line). 
The calculations have been performed for couplings 
 $J=2$ and $h=1$, and for $N=10$ particles. At low temperatures,
 we find $\langle H_J \rangle_T/N \approx -J$ as expected. 

%%%%%%%%%%%%%%%%%%%%%%%%%%%%%%%%%%%%
%%%%%%%%%%%%%%%%%%%%%%%%%%%%%%%%%%%%

Alternatively, if we want to calculate the thermal average of the 
one-body interaction, $H_h = -h\sum_i s_i$, we need to define,
\begin{equation}
H^{\lambda_2} = -J  \sum_{\langle i,j\rangle} s_i s_j 
- \lambda_2 h \sum_i s_i \,.
\label{eq:i8}
\end{equation}
The corresponding partition function $Z^{\lambda_2}$ becomes
\begin{align}
        Z^{\lambda_2}= e&^{N \beta J} \left ( \left [ \cosh \beta \lambda_2 h 
        + \sqrt { \sinh^2 \beta \lambda_2 h 
        + e^{- 4  \beta J}} \right ]^N \right . \nonumber \\ 
       & + \left .  \left [ \cosh \beta \lambda_2  h 
       - \sqrt { \sinh^2 \beta \lambda_2 h 
       + e^{- 4  \beta J }} \right ]^N \right ) \, .
\label{eq:i9}
\end{align}
The thermal average of $H_h$ is then calculated as
\begin{equation}
\langle H_h \rangle_T = \left.
 \frac {\partial F^{\lambda_2}}{\partial \lambda_2}\right|_{\lambda_2 = 1} 
 \, .
\label{eq:i10}
\end{equation}
$\langle H_h \rangle_T/N$ is shown as a function of temperature
in Fig.~~\ref{fig:ising} (dashed line). As expected, as $T \to 0$, 
we find that $\langle H_h \rangle_T/N \to -h$. In contrast, in the classical
limit, the temperature washes out any residual magnetisation and 
$\langle H_h \rangle_T$ asymptotically tends to $0$.

We also show in Fig.~\ref{fig:ising} the sum of the two terms in 
the Hamiltonian (dotted line). We compare this sum with an explicit
calculation of the thermal average of the full Hamiltonian in Eq.~(\ref{eq:i1}) 
(circles). We find a perfect agreement within numerical errors. 
Notice that at zero temperature, the expectation 
value of the full Hamiltonian is $\langle H \rangle_T/N = -(J+h)$. In other
words, the $H_J$ and $H_h$ terms contribute to the total energy per particle 
in proportion to their respective couplings $J$ and $h$.
In contrast, at very high temperatures, $T \gg h$ and $T \gg J$, the thermal average of the Hamiltonian decays with temperature,
$\langle H \rangle_T/N = -(J^2+h^2)/T$. The two contributions now 
scale with $J^2$ and $h^2$. In the specific example shown in Fig.~\ref{fig:ising}, the $J$ term dominates at high temperatures by a factor of 4. 

The HF theorem for the Ising model allows us to compute separately the thermal 
expectation values of $H_J$ and $H_h$. From a practical perspective, this provides an alternative way to 
compute the total energy of the system. But, more importantly, 
the HF theorem provides access to
contributions of the energy which are associated to physical properties, like the degree of 
magnetization. From a physics perspective, these contributions carry relevant information 
on the system's structure. 

%%%%%%%%%%%%%%%%%%%%%%%%%%%%%%%%%%%%%%%%%%%%%%%%%

\section{Lipkin model}
\label{sec1}

The last example will be devoted to a model proposed in the mid-60s of 
the last century by Lipkin, Meshkov and Glick to describe $N$ 
fermions occupying two energy levels, with each level having an 
$N$-fold degeneracy.~\cite{lipkin1965} This model has been used in 
the context  of nuclear physics and also as a laboratory to test 
many-body techniques.~\cite{kuo1994,co2018} From a quantum mechanics
teaching perspective, this example 
illustrates the use of creation and annihilation operators; the 
construction of Fock spaces; and the numerical diagonalization of 
a finite size Hamiltonian. We note that this model has also been explored at finite temperature 
in the past. \cite{Feng1979,Rossignoli1984,Wei2018,Pons2018}

The  Hamiltonian\footnote{From now on, we explicitly display a hat, $\hat H$, in the Hamiltonians to stress the fact that these are second-quantised operators.} of this model is typically made up of three terms:
\beqa
\hat H = \varepsilon \hat J_0 - \frac {V}{2} (\hat J_+^2 + \hat J_-^2 ) 
- \frac {W}{2} (\hat J_+ \hat J_- + \hat J_- \hat J_+)  \, ,
\label{eq:l1}
\eeqa
with 
\beqa
\hat J_0 &=& \frac {1}{2} \sum_{p} 
\left [ \hat a_{p,1}^{\dagger}  \hat  a_{p,1} - \hat  a_{p,-1}^{\dagger} \hat a_{p,-1} \right ] 
\,,\nonumber \\
\hat J_+ &=& \sum_p \hat a_{p,1}^{\dagger} \hat a_{p,-1}~,~~ 
\hat J_- = \sum_{p} \hat a_{p,-1}^{\dagger} \hat a_{p,1}  \, ,
\label{eq:l2}
\eeqa
where $\hat  a_{p,\pm 1}^{\dagger}$ and $\hat  a_{p,\pm 1}$ are the creation and
annihilation operators on the top ($+1$) or bottom ($-1$) level in site $p=1,\cdots,N$.
These operators satisfy the anti-commutation relations,
$ \{ \hat  a_{p,\alpha}, \hat a_{r,\beta}^{\dagger}\} = 
\delta_{p,r} \delta_{\alpha,\beta}$ and 
 $ \{ \hat  a_{p,\alpha}, \hat a_{r,\beta}\}= \{ \hat a_{p,\alpha}^{\dagger}, \hat a_{r,\beta}^{\dagger}\}=0.$ 

The first term in the Hamiltonian is associated to $\hat J_0$ and describes the 
system in absence of interactions. 
$\varepsilon$ is the  energy difference between the two levels, and is used
as an energy unit in the reminder of this paper. 
The Fock 
states built with the creation operators diagonalize $\hat J_0$ 
and the total energy is proportional to the difference between the number 
of particles in the top and bottom levels. The dimension of this 
Fock space is $2^N$ as, in each of the $N$ sites, particles can either be in 
the top or the bottom state, with one particle in each site. 
Any isolated two-level system is isomorphous to a spin$-1/2$ system, and so
the operators $\hat J_0, \hat J_+$ and $\hat J_-$ satisfy 
commutation relations that are characteristic of the algebra of angular 
momentum,
\begin{equation}
	\left [\hat J_+,\hat J_- \right ] = 2 \hat J_0~,
	~\left [ \hat J_0 , \hat J_+\right ] = \hat J_+~,~
	\left [ \hat J_0 , \hat J_- \right ] = - \hat J_- \, .
\end{equation}
One can then introduce the Casimir operator 
$\hat J^2 = \frac{1}{2} \left( \hat J_+ \hat J_- + \hat J_- \hat J_+ \right) + \hat J_0^2 $ 
and 
make use of the analogy with the angular momentum algebra to define 
the action of these operators on eigenvectors defined by the quantum numbers $j$ and $m$, $\mid j,m\rangle$,
\begin{align}
	\hat J_0 \mid j,m\rangle &= m \mid j, m\rangle \,, \nonumber \\
	\hat J^2 \mid j,m \rangle & =  j(j+1) \mid j,m \rangle \, , \nonumber \\ 
	\hat J_{\pm} \mid j,m \rangle &= \sqrt {j(j+1) -m(m\pm 1)}  
	\mid j, m\pm1 \rangle \, .
	\label{eq:l4}
\end{align}
These eigenvectors are linear combinations of the Fock states constructed by acting 
with the creation operators on the vacuum.  
The matrix elements of the Hamiltonian are readily calculated using Eq.~({\ref{eq:l4}). 
For our purposes, {\it i.e.} to illustrate the fulfillment of the 
HF theorem, we neglect
the contributions due to $W$ and  consider only the interaction term proportional to $V$. 
There are only
three non-zero combinations of matrix elements for a given value of $j$,
\begin{align}
\langle j,m \mid \hat H \mid j, m+2 \rangle =& 
-\frac {V}{2} \sqrt {\left (j (j+1) - m(m+1) \right ) } \nonumber \\
&\times \sqrt {\left ( j (j+1) - (m+1)(m+2) \right ) } \, ,\nonumber \\
\langle j, m \mid \hat H \mid j,m \rangle &= \varepsilon \,m \, , \nonumber \\
\langle j, m+2 \mid \hat H \mid j, m \rangle &=  \langle j, m \mid \hat H \mid j, m+2 \rangle \, .
\end{align}

At zero temperature, it is customary to set $j=N/2$,  since the 
ground state of the system lies in the subspace with the maximum value 
of $j$. 
$m$ takes the values $m=-j,-j+1, \cdots, ,j-1,j$, and for a given $j$ this sets a subspace of dimension $2 j+1$. 
In a finite temperature setting, thermal
fluctuations populate the states with $j < \frac{N}{2}$. 
For an even (odd) number of particles $N$, $j$ can therefore take the values $j=j_\text{min},\cdots,N/2$ with $j_\text{min}=0$ ($j_\text{min}=1/2$). 

The diagonalization of  the Hamiltonian provides 
the eigenenergies of the different subspaces, $E_k^{(j)}$, with $k=1, \cdots, 2j+1$. The ground state is non degenerate, but
different subspaces with the same value of $j$ generate the same eigenenergies.
The multiplicity
\begin{equation}
\alpha^N_j = \frac {1 + 2 j}{1 + j + \frac {N}{2}} 
\binom {N}{\frac {N}{2} - j}, 
\label{eq:l6}
\end{equation}	
counts the number of times that an angular momentum $j$ occurs for a given $N$, i.e. the number of different SU(2)  subspaces of dimension $2j+1$ arising from coupling $N$ spins. \cite{Feng1979,Rossignoli1984}
Adding up the multiplicities of each subspace in $j$, we recover the total dimension of the Fock space,  $\sum_j (2j+1) \alpha^N_j = 2^N$. 

We define the parametric Hamiltonian $\hat H^\lambda$ as 
\begin{equation}
\hat H^\lambda = \varepsilon \hat J_0 
- \lambda \frac {V}{2} (\hat J_+^2 + \hat J_-^2) \, ,
\end{equation}
so that $\hat H_1=\frac {V}{2} (\hat J_+^2 + \hat J_-^2)$.
After diagonalizing  $\hat H^\lambda$ in each of the 
$j-$subspaces, one obtains the eigenenergies $E_k^{(j),\lambda}$ with
$k=1, \cdots, 2j+1$.
At finite temperature, the partition function is numerically calculated 
from the sum
\begin{equation}
Z^\lambda = \sum_{j=j_\text{min}}^{N/2} \alpha^N_j \sum_{k=1}^{2j+1} 
e^{- \beta E_k^{(j),\lambda}} \, .
\end{equation}
We can then calculate the energy, Eq.~(\ref{eq:ener}), and the free energy, $F^\lambda= - T \ln Z^\lambda$.
The thermal average of the interaction energy is then given by the HF theorem at finite temperature, Eq.~(\ref{eq:HFt}).
Alternatively, one can explicitly compute the expectation value of the interaction energy 
in each eigenstate, and perform the thermal average separately,
\begin{align}
\langle H_1 \rangle_T &= \frac {1}{Z} \sum_{j=j_\text{min}}^{N/2} \alpha^N_j \sum_{k=1}^{2j+1} 
\langle j, k | \hat  H_1 | j, k \rangle 
 e^{-\beta E_k^{(j)}} \, .
\label{eq:avexp}
\end{align}

%%%%%%%%%%%%%%%%%%%%%%%%%%%%%%%%%%%%
\begin{figure}[t!]
\centering
\includegraphics[width=0.75\columnwidth]{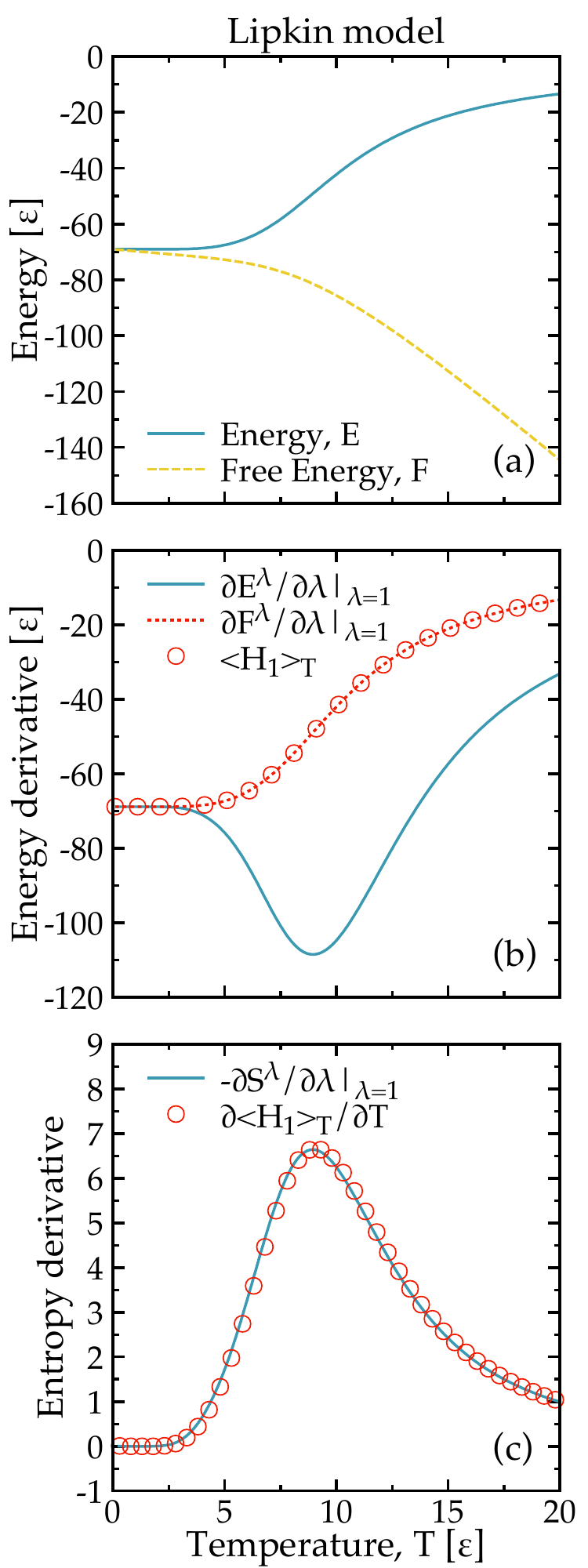}
\caption{(a) Energy $E$ (solid line) and free energy $F$ (dashed line) 
as a function of temperature for the Lipkin model in units of $\varepsilon$. 
(b) Derivative  of $E^\lambda$  (solid line) and $F^\lambda$ (dotted line) with respect to $\lambda$ at $\lambda=1$ as a function of temperature. 
(c) Minus the derivative of $S^\lambda$  with respect to $\lambda$ at $\lambda=1$ 
as a function of temperature. Circles correspond to the temperature derivative $\frac {\partial  \left\langle H_1 \right\rangle_T}{\partial T }$. All calculations in this figure correspond to $N=10$ and $V=3$.
}
\label{fig:lipkin}
\end{figure}

The three panels in Fig.~(\ref{fig:lipkin}) provide an example of the results 
for the thermal Lipkin model. We choose $N=10$ particles and $V=3$ as an 
illustrative test. We report the total energy (solid line) and the free energy
(dashed line) as a function of temperature
in panel (a). Both $E$ and $F$ coincide at $T=0$. As temperature increases,
however, the energy is an increasing function of temperature. In contrast,
the free energy decreases with temperature due to the 
contribution of the entropy.

The fulfillment of the HF theorem is illustrated in panel (b). 
We show $\left. \partial E^\lambda/\partial \lambda  \right|_{\lambda = 1} $ 
(solid line) and $\left. \partial F^\lambda / \partial \lambda \right|_{\lambda = 1} $
(dotted line). The empty circles show the thermal average of 
the interaction term, $\langle \hat H_1 \rangle_T$. We stress that this 
average has been explicitly calculated using Eq.~(\ref{eq:avexp}) rather
than the derivative of $F^\lambda$ with respect to $\lambda$.
We find a perfect agreement between these two independent calculations,
thus providing a numerical proof of the generalization of the HF theorem to
finite temperature. 
We note that for this example with $V=3$, $E$ in panel (a) and $\langle H_1 \rangle_T$ in panel (b)
are very similar.

The solid line  in panel (b) corresponds to the energy derivative $\left. \partial E^\lambda / \partial \lambda \right|_{\lambda = 1} $. 
The differences between the solid and dotted curves indicate the importance of 
the entropy contribution to the derivative of $F^\lambda$ with 
respect to $\lambda$.  At  low temperatures, $T \lesssim 4$, the differences 
between the derivatives of $F^\lambda$ and $E^\lambda$ are very small.
At some point when increasing temperature, though, the energy derivative shows a dip with a minimum around
$T \approx 10$, and a subsequent increase with temperature. In contrast, the
derivative of the free energy (and the thermal average of $\hat H_1$) are increasing functions
of temperature throughout. 

Finally, panel (c) of Fig.~\ref{fig:lipkin} shows the relation between the derivative 
of $S^\lambda$ with respect to $\lambda$ (solid line) and 
$ \partial \langle \hat H_1 \rangle / \partial T $ (empty circles). This derivative is directly 
related to the slope of $\langle \hat H_1 \rangle_T$ as a function of temperature, shown in panel (b). 
At low temperatures, the temperature dependence of $\langle H_1 \rangle_T$ is 
mild and the derivative is close to zero. At intermediate temperatures,
the entropy derivative shows a maximum, in the same region where the
energy derivative in panel (b) has a minimum. 
For $T \rightarrow \infty$,  since the size of 
the Hilbert space we are considering is finite, the system tends 
to equipartition, {\it i.e.,} all states are equally populated. In particular,
the entropy of the system in this limit is $S \to \ln 2^N$, and one expects it to 
be temperature (and $\lambda$) independent. \cite{Pons2018} Similar arguments hold for the energy and 
the free energy, so their derivatives saturate and cancel 
at (classically) high temperatures. The slow decrease of the derivative in panel (c) confirms
numerically these expectations.

\section{Summary and conclusions}

The main purpose of this paper is to present a simple derivation
of the Hellmann-Feynman theorem at finite temperature. Our aim
is to illustrate its usefulness as a pedagogical tool in teaching
quantum mechanics at finite temperature. We note that previous work 
on the subject has focused on specific research 
questions.~\cite{cabrera1989,fan1995,Popov1998,rai2007}

The theorem allows one
to calculate the thermal average of operators contained in the
Hamiltonian, by calculating 
the derivative of the free energy with respect to a 
parameter that modulates the action of operators in the 
Hamiltonian. The theorem  is specially useful for cases where 
one has access to the free energy, but not necessarily to the 
separate expectation values of the individual terms in the Hamiltonian, like the 
kinetic and/or the interaction terms. 
We have also derived a relation 
between the variation of the entropy with respect to this 
parameter and the temperature derivative of the thermal 
average of the operator. This relation is not trivial \emph{a priori}. 

We have illustrated the usefulness of the HF theorem at finite temperature
with three different examples that can be used in quantum 
mechanics or quantum statistical mechanics courses. 
Two of these examples, concerning the one-dimensional harmonic 
oscillator and the Ising model, are fully analytical and can be exploited
in undergraduate courses. 
The third 
example, the Lipkin model, requires some knowledge of 
second quantization and the use of numerical 
diagonalization techniques, so it may be useful for 
more advanced, master-level courses. 

%%%%%%%%%%%%%%%%%%%%%%%%%%%%%%%%%%%%%%%%%%%%%%%%%
\begin{acknowledgments}
We thank Aurel Bulgac and Pedro P{\'e}rez-Fern{\'a}ndez for useful correspondance.
We acknowledge  financial support from the Spanish Ministerio de 
Economia y Competitividad Grant No FIS2017-87534-P, from 
Generalitat de Catalunya Grant No. 2014SGR401, the COST Action
CA 16214, and the UK Science and Technology Facilities Council (STFC) grant ST/P005314/1.
\end{acknowledgments}

%%%%%%%%%%%%%%%%%%%%%%%%%%%%%%%%%%%%%%%%%%%%%%%%%

%%%%%%%%%%%%%%%%%%%%%%%%%%%%%%%%%%%%%%%%%%%%%%%%%%%%%%%%%%%%%%%%%%%%%%%%%%%%%%%%%%%%%%%%%%
% BIBLIOGRAPHY

\end{document}